\documentclass[12pt]{article}
\usepackage{amsfonts,amsmath,amssymb}

\def\zz{{\mathbb Z}}
\def\rr{{\mathbb R}}
\def\nn{{\mathbb N}}
\def\ff{{\mathbb F}}
\def\ul{\underline}
\def\ba{\begin{array}}
\def\ea{\end{array}}
\def\bt{\begin{tabular}}
\def\et{\end{tabular}}
\def\hy{{\hspace*{+3 mm}}}
\def\ds{\displaystyle}
\begin{document}

\begin{centering}

{

\large{\bf The Battery-Discharge--Model:}

\large{\bf
A Class of Stochastic Finite Automata to Simulate
}

\large{\bf
Multidimensional Continued Fraction Expansion
}

}
\vspace*{ 0.5 cm}

Michael Vielhaber\footnote{Supported  by FONDECYT 1040975. Partly
  supported by DID UACh.}
 and M\'onica del Pilar Canales Ch.\footnotemark[\value{footnote}]

\vspace*{ 0.5 cm}

Instituto de Matem\'aticas\\
Universidad Austral de Chile\\
Casilla 567,Valdivia

{\tt vielhaber@gmail.com\ \ \ monicadelpilar@gmail.com}

\end{centering}

\vspace*{ 0.5 cm}

{\sc Abstract}
We define an infinite  stochastic  state machine, the 
{\it Battery--Discharge--Model} (BDM), which simulates the  
behaviour of linear and jump complexity of the  continued fraction
expansion of multidimensional 
formal power series, a relevant security measure in the
cryptanalysis of stream ciphers.

We also obtain finite approximations to the infinite BDM, where {\it
  polynomially} many states suffice to approximate with an 
{\it  exponentially small} error  the  probabilities and averages  for 
linear and jump  complexity of $M$--multisequences of length $n$ over
  the finite field $\ff_q$, for {\it any}  $M$, $n$,  $q$.

\subsubsection*{Introduction}

In cryptography two important measures of sequence complexity are the
linear and jump complexity, dealing with the  
continued fraction expansion of the sequence seen as 
formal power series  over some finite field $\ff_q$.
While both complexities are well understood for single sequences
(\cite{NV},\cite{V}), a current topic is to generalize these notions
to multisequences ($M$ streams of symbols in parallel) with first
results for $M=2,q=2$ given in \cite{Dai}.

In Section I we suggest an infinite recurrent stochastic automaton and
finite approximations, the {\it Battery--Discharge--Model } that 
simulates the continued fraction expansion (proof in Section II) and thus (Section III)
answers questions about linear and jump complexity 
for every $M$ (``multi''--ness), $q$ (order of finite field), and $n$
(length of  sequence).
\subsection*{I. The Battery--Discharge--Model}

In this first part, we develop in three steps an  infinite  stochastic
automaton, the  {\it Battery--Discharge--Model} and a family of finite
approximations.

\subsubsection*{a) Model without discharge}

We fix a number  $M\in \nn$  and then have $M$ batteries, 
each holding a charge $b_i\in\zz, 1\leq i \leq M$, 
and a drain $d\in\zz$, hence an infinite number of possible
states. 
The  initial state is  $d = b_i = 0$.

The model cyclically runs through $M+1$ main cycles $T=0,\dots,M$.
At each transition $T\to T+1$ for $T=0,\dots , M-1$, the drain is decremented, 
$d := d-1$, whereas the batteries do not change. 
At the transition from $T=M$ to $T=0$, all batteries are incremented, 
$b_i := b_i + 1, 1\leq i \leq M$, whereas $d$ remains constant. With the 
initial condition $d=b_i=0$, we thus have
$$T + d + \sum_{i=1}^M b_i \equiv 0\mbox{\rm\ \ \ (invariant)} $$
Writing the state in the form $(b_1,\dots,b_M;d)_T$, we obtain the 
following behaviour for this model:\ \ \ 
$(0,\dots,0;0)_0 \to(0,\dots,0;-1)_1 \to(0,\dots,0;-2)_2\to\dots
\to(0,\dots,0;-M)_M \to(1,\dots,1;-M)_0 
\to(1,\dots,1;-M-1)_1\to$\\
$(1,\dots,1;-M-2)_2\to\dots \to(1,\dots,1;-2M)_M 
\to(2,\dots,2;-2M)_0 \to\dots$

\subsubsection*{b) Model with discharge}

Each of the $M+1$ major cycles now is divided into $M+1$ subcycles $t=*,1,\dots,M$.
Subcycle $*$ shows the result of decrementing $d$ or incrementing the
$b_i$, whereas 
during subcycle $t, t=1,\dots,M$, battery $b_t$ may {\it discharge} into the drain, 
provided it has high enough potential that is $b_t > d$. In this case the 
excess charge goes from $b_t$ to the drain, amounting to an interchange 
$d \leftrightarrow b_t$ of values, thus maintaining the invariant.

The behaviour with discharge is as follows (for illustration we use $M=3$):
The underlined battery is the one, $b_t$, corresponding to the subcycle.
We show the result at the {\it end} of the subcycle. 
In case of a discharge, (the new) $b_t$  is in boldface:

{\small{
$(b_1,b_2,b_3;d)_{T,t} = 
(0,0,0;0)_{0,*}\to
(\ul{0},0,0;0)_{0,1}\to
(0,\ul{0},0;0)_{0,2}\to
(0,0,\ul{0};0)_{0,3}\to\\
(0,0,0;-1)_{1,*}\to
({\ul{\bf -1}},0,0;0)_{1,1}\to  
(-1,{\ul{0}},0;0)_{1,2}\to
(-1,0,{\ul{0}};0)_{1,3}\to\\  
(-1,0,0;-1)_{2,*}\to  
({\ul{-1}},0,0;-1)_{2,1}\to  
(-1,{\ul{\bf -1}},0;0)_{2,2}\to  
(-1,-1,{\ul{0}};0)_{2,3}\to\\
(-1,-1,0;-1)_{3,*}\to  
({\ul{-1}},-1,0;-1)_{3,1}\to  
(-1,{\ul{ -1}},0;-1)_{3,2}\to  
(-1,-1,{\ul{\bf -1}};0)_{3,3}\to\\
(0,0,0;0)_{0,*}\to
(\ul{0},0,0;0)_{0,1}\to
(0,\ul{0},0;0)_{0,2}\to
(0,0,\ul{0};0)_{0,3}\to\dots$
}}

The behaviour of this model I.b) is  (purely) periodic.

\subsubsection*{c) Model with discharge and inhibition}

Finally we introduce the stochastic element of temporary inhibition of 
a battery (observe that in I.a) the batteries never discharge, in I.b) always):

Let $\frac{1}{q}$, with $0 < \frac{1}{q}\leq 1$, be the probability that
a battery with $b_t>d$ will be {\it inhibited} and thus will not discharge.
We now have a stochastic behaviour:
$(0,0,0;0)_{0,*} \to \dots (0,0,0;-1)_{1,*}$ with probability 1, but
$(0,0,0;-1)_{1,*} \to ({\ul\bf{-1}},0,0;0)_{1,1}$ with probability $\frac{q-1}{q}$, and
$(0,0,0;-1)_{1,*} \to ({\ul{0}},0,0;-1)_{1,1}$ with probability
$\frac{1}{q}$, and then  battery 2 may be inhibited or not etc. 

A more involved example: From $(0,2,1;-4)_{1,*}$ within cycle $T=1$,
six outcomes are possible
(``$D$'' = discharge, ``$I$'' = inhibition, ``$-$'' = $b_i \leq d$):

$
\ba{cccccccc}
(b_1,b_2,b_3;d)_{1,*}&b_1&b_2&b_3&Prob&(d;b_1,b_2,b_3)_{1,3}&\to&(d;b_1,b_2,b_3)_{2,*}\\
(0,2,1;-4)&D&D&-&(q-1)^2 / q^2&(-4,0,1;2)&\to&(-4,0,1;1)\\
(0,2,1;-4)&D&I&D&(q-1)^2 / q^3&(-4,2,0;1)&\to&(-4,2,0;0)\\
(0,2,1;-4)&D&I&I&(q-1)   / q^3&(-4,2,1;0)&\to&(-4,2,1;-1)\\
(0,2,1;-4)&I&D&-&(q-1)   / q^2&(0,-4,1;2)&\to&(0,-4,1;1)\\
(0,2,1;-4)&I&I&D&(q-1)   / q^3&(0,2,-4;1)&\to&(0,2,-4;0)\\
(0,2,1;-4)&I&I&I&    1   / q^3&(0,2,1;-4)&\to&(0,2,1;-5)\\
\ea
$

For instance, the transition of the second line consists of these steps:

{\small{$
(0,2,1;-4)\stackrel{D;\frac{q-1}{q}}{\longrightarrow}
({\ul{\bf{-4}}},2,1;0)\stackrel{I;\frac{1}{q}}{\longrightarrow}
(-4,\ul{2},1;0)\stackrel{D;\frac{q-1}{q}}{\longrightarrow}
(-4,2,\ul{\bf{0}};1)\stackrel{d:=d-1}{\longrightarrow}
(-4,2,0;0)$}})

\subsubsection*{d) Properties of the full model\footnote{..which of
    course could be called the Butterfly-Dinner-Model, where $M$
    flowers $f_i$ are visited in turn by a butterfly, dining from the
    supplied nectar, whenever the level exceeds that of the butterfly,
    with probability $\frac{1}{q}$ that the flower has its petals closed...}}

This is our full model.  
We shall use only the states at timesteps $(T,*)$ and let the
transition probability take care of the events (discharge, inhibition) during 
subcycles $t=1,\dots,M$.
The state set is isomorphic to 
$\zz^M\times \{0,\dots,M\}$, where the $M$ battery levels and 
the main cycle are given and the drain value results implicitly 
from $d = -T -\sum_{i=1}^M b_i$.

Every transition probability is  of the form
$\frac{(q-1)^a}{q^b}$  for $b$ batteries with $b_i>d$ at their
subcycle, with $a$ discharges and $b-a$ inhibitions.
So, all transition  (and state) probabilities are polynomial functions
in $q$ by construction.
The values for $b_i,d$ in the successor state $(T+1,*)$ already
incorporate the decrement of $d$ or increment of the $b_i$.

Let $Q_T$ be the (infinite) set of all states in main cycle
$T,T=0,\dots,M$. 
We adjoin a  class $K\in\nn_0$ to each state as follows:
State $(0,\dots,0;0)_{0,*}$ and all states reachable without inhibition from here 
are in class 0 (these are just the states $(\dots)_{T,*}$ of model I.b). 
All states reachable from a class $K$ with 
$i\in\nn_0$ inhibitions belong to  class $K+i$. If a state can be reached 
in different ways (number of inhibitions), 
the {\it smallest} such class number applies.
The rationale for these classes is:

$i$) Numerical evidence shows that in the stationary distribution of
the infinite model in each cycle $T$, {\it every} state of  
class $K$ occurs with   probability 
{\it exactly} $q^{-K}$ times the probability of the (unique) state  in
class 0. 

$ii$) For $q\to\infty$, almost always we are in 
the states of class 0.
Hence, restricting the (infinite) state set $\zz^M\times \{0,\dots,M\}$ to 
those states within classes $0\dots K_0$, for some fixed accuracy 
$K_0\in\nn_0$, we obtain -- at least for large $q$ -- a fairly good
approximation to the infinite model. 
In this case, for a state in class $K$, the model allows  at most 
$K_0 - K$ more  inhibitions and 
thus the $(K_0-K+1)$-st battery with $b > d$ {\it has} to discharge (with 
probability 1).

$iii)$ Also, simulation results for $K_0$ up to 120 show that the
number of states in class $K$ for the unbounded model and for each
main cycle $T$ is $p_K(M)$,  the number
of partitions of $K$ into at most $M$ parts or -- what is the same -- 
into parts of size at most $M$. $p_K(M)$ grows as
$\approx\frac{K^{M-1}}{M!(M-1)!}$ asymptotically for fixed $M$ and
$K\to\infty$ 
(see INRIA \cite{INRIA} and Sloane's \cite{NJAS} integer sequences).
Thus the bounded finite models have an overall number of states 
$(M+1)\cdot \sum_{K=0}^{K_0} p_K(M)$.
Furthermore, let us define 
${\cal P}(M,q)=\sum_{K\in\nn_0}p_K(M)\cdot q^{-K}=\sum_{s\in Q_T}
q^{-K(s)}$ (the same for every  $T$). Then a state of class $K$ has
probability $q^{-K}/{\cal P}(M,q)$.

{\bf Example}\ \ 
The bounded model for $M=3$ and $K_0=2$ consists of $(3+1)\cdot
\sum_{K=0}^2p_K(3)=16$  states (one state  each in class 0 and 1, two
states in class 2, for each $T$).
Since ${\cal P}(3,q)=q^3/((q-1)^2(q+1))$, already this {\it very}
limited model accounts for a share $(1\cdot q^0+1\cdot q^1+2\cdot
q^2)/{\cal P}(M,q)$ of the stationary probability distribution of the
unbounded model, which is $75\%$ for $q=2$, $99.6\%$ for $q=8$ etc.

Here we give all states with class $K\leq K_0=2$, belonging to
subcycle $(T,*)$ together with all of their successor states in
subcycle $(T+1,*)$ and the respective transition probability. In the
first line {\it e.g.}, we have a probability of $1/q^2$ to go to
$(0,0,-1;0)$ instead of $(q-1)/q^3$ (for 2 inhibitions and 1
discharge), since battery 2 {\it has} to discharge to keep within
$K\leq K_0$, state $(0,0,0;-1)$ does not appear in this bounded model.

{\small{
$\ba{cc|c|l}
T&K&(b_1,b_2,b_3;d)&\mbox{\rm Nextstates\ :\ \ Probability}\\
0&0&(0,0,0;0)&(-1,0,0;0):\frac{q-1}{q},\ \ (0,-1,0;0):\frac{q-1}{q^2},\ \ (0,0,-1;0):\frac{1}{q^2}\\
0&1&(-1,0,0;1)&(-1,0,0;0):1\\
0&2&(0,-1,0;1)&(0,-1,0;0):1\\
0&2&(-1,0,1;0)&(-1,-1,0;1):1\\
1&0&(-1,0,0;0)&(-1,-1,0;0):\frac{q-1}{q},\ \ (-1,0,-1;0):\frac{q-1}{q^2}\
(-1,0,0;-1):\frac{1}{q^2}\\
1&1&(0,-1,0;0)&(-1,-1,0;0):\frac{q-1}{q},\ \ (0,-1,-1;0):\frac{1}{q}\\
1&2&(0,0,-1;0)&(-1,0,-1;0):\frac{q-1}{q},\ \ (0,-1,-1;0):\frac{1}{q}\\
1&2&(-1,-1,0;1)&(-1,-1,0;0):1\\
2&0&(-1,-1,0;0)&(-1,-1,-1;0):\frac{q-1}{q},\ \ (-1,-1,0;-1):\frac{1}{q}\\
2&1&(-1,0,-1;0)&(-1,-1,-1;0):\frac{q-1}{q},\ \ (-1,0,-1;-1):\frac{1}{q}\\
2&2&(-1,0,0;-1)&(-2,-1,0;0):1\\
2&2&(0,-1,-1;0)&(-1,-1,-1;0):1\\
3&0&(-1,-1,-1;0)&(0,0,0;0):1\\
3&1&(-1,-1,0;-1)&(-1,0,0;1):\frac{(q-1)^2}{q^2}\
(-1,0,1;0):\frac{q-1}{q^2},\ \ (0,-1,0;1):
\frac{1}{q}\\
3&2&(-1,0,-1;-1)&(-1,0,0;1):\frac{q-1}{q},\ \ (0,-1,0;1):\frac{1}{q}\\
3&2&(-2,-1,0;0)&(-1,0,0;1):\frac{q-1}{q},\ \ (-1,0,1;0):\frac{1}{q}\\
\ea$
}}

\subsection*{II. The BDM and Continued Fraction Expansion}

We now apply the BDM to obtain precise values about the behaviour of
the 
linear and jump complexity of multisequences: Let
$G_t(a)=\sum_{i=1}^\infty a_{t,i}x^{-i}\in\ff_q[[x^{-1}]],t=1,\dots,M$
be $M$ formal power series over the finite field $\ff_q$.

The linear complexity of $(G_t(a)\ | \ 1\leq t\leq M)$ at $n$ is
defined as the smallest degree of a polynomial $v(x)$, such that there
are some polynomials $u_t(x), 1\leq t\leq M$ with $G_t(x)
=\frac{u_t(x)}{v(x)}+O(x^{-(n+1)})$. The jump complexity in turn counts,
how often this smallest degree has changed (increased) until step
$n$ (see \cite{NVsim}\cite{VSETA}).

We derive these complexities from our BDM, using its equivalence to
the multi-Strict Continued Fraction Algorithm (m--SCFA) of Dai and
Feng \cite{Dai}.

The m--SCFA uses the following variables to describe the state:

$n$, the timestep

$d =: d_{\text{SCFA}}$, the degree of $v$, the current approximation denominator

$w_t, 1\leq t\leq M$, a ``degree deviation'' of $u_t(x)$ at sequence
$t$

Our BDM uses the equivalent variables:

$(*)$ $T$, timestep, with $T\equiv n$ mod $(M+1)$

$(**)$ $d=: d_{\text{BDM}}$, $d_{\text{BDM}} = d_{\text{SCFA}}
-\left\lceil\frac{n\cdot M}{M+1}\right\rceil$, the deviation of
$deg(v)$ from its typical value,

$(***)$ $b_t$, $b_t= \left\lfloor\frac{n}{M+1}\right\rfloor - w_t,
1\leq t\leq M$, the battery levels. 

Observe that initially (at $n=T=0$) $d_{\text{SCFA}} = d_{\text{BDM}}
= w_t=b_t=0,\forall t$, so both models coincide according to
equivalences $(*)$ to $(***)$.

Let us first consider the timestep $n$, main cycle $T$, at subcycle $*$: \\
Assuming $d_\text{SCFA},w_t$ fix with $n\to n+1$ we must have the new
values
$$d^+_\text{BDM}\stackrel{(**)!}{=}
d_\text{SCFA} -\left\lceil\frac{(n+1)\cdot M}{M+1}\right\rceil=
d_\text{SCFA} -\left\lceil\frac{n\cdot M}{M+1}\right\rceil-\varepsilon=
d_\text{BDM}-\varepsilon$$
where $\varepsilon=0$ for $n+1\equiv 0\mod (M+1)$ and 1 otherwise, and
$$b^+_t\stackrel{(***)!}{=}
\left\lfloor\frac{n+1}{M+1}\right\rfloor - w_t = 
\left\lfloor\frac{n}{M+1}\right\rfloor +\varepsilon - w_t = 
b_t+\varepsilon,\forall t$$
where $\varepsilon=1$ for $n+1\equiv 0\mod (M+1)$ and 0 otherwise.
This corresponds to incrementing the $b_i's$  for $T\equiv M\to 0$ and
otherwise decrementing $d$.

Now, within the $M$ subcycles $t=1,\dots,M$ we consider four cases,
according to a ``discrepancy'' $\delta$ of the m--SCFA (the deviation
between the formal power series and the approximation by $u_t(x)/v(x)$) and the
values of $n,d,w_t$:

\bt{cccl}
 & m--SCFA &\cite[Thm.~2]{Dai}& BDM Case\\
1&$\delta = 0$    and $n-d_\text{SCFA} \leq w_t$ &a&level too low, ``--''\\
2&$\delta \neq 0$ and $n-d_\text{SCFA} \leq w_t$ &c&level too low, ``--''\\
3&$\delta = 0$    and $n-d_\text{SCFA} >    w_t$ &a&inhibition ``I''\\
4&$\delta \neq 0$ and $n-d_\text{SCFA} >    w_t$ &b&discharge ``D''\\
\et

First note that 
$n-d_\text{SCFA} > w_t \Leftrightarrow n-\left(\left\lceil 
\frac{n\cdot M}{M+1}\right\rceil+ d_\text{BDM}\right) 
> \left\lfloor \frac{n}{M+1}\right\rfloor -b_t \Leftrightarrow b_t >
d_\text{BDM}$ corresponds to cases 3 and 4, that is discharge or
inhibition. We model a discrepancy 
value $\delta = 0$ by the probability of inhibition $1/q$, according
to the following proposition about the even distribution of
discrepancy values.

{\bf Proposition}\ \ 
{\it
In any given position $(m,n),  1\leq m\leq M, n\in\nn$ of the formal
power series, 
exactly one choice for the next symbol  $a_{m,n}$
will yield a dis\-cre\-pan\-cy $\delta=0$, all other $q-1$ symbols from $\ff_q$ 
result in some  $\delta\neq 0$.
}

{\bf Proof:}\ 
The current approximation $u_m^{(m,n)}(x)/v^{(m,n)}(x)$ determines
exactly  {\it one } approximating coefficient sequence  for the $m$--th
formal power series $G_m$. 
The (only) corresponding symbol belongs to $\delta=0$.
\hfill $\Box$ 

In fact, for every position 
$(m,n)$, each discrepancy value $\delta \in\ff_q$
occurs exactly once for some  $a_{m,n}\in\ff_q$, 
in other words (compare \cite{CV}\cite{VSETA} for $M=1$):

\vspace{5 mm}

{\bf Fact}
\ \ 
{\it The Generalized Berlekamp--Massey--Algorithm $($GBMA$)$
and the multi--Strict Continued Fraction Algorithm $($sCFA$)$ 
induce an isometry on $(\ff_q^M)^\omega$.
}

Concerning the update of the $d_\text{SCFA}$ and $w_t$ values, in
cases 1 to 3 nothing happens, neither in the m--SCFA, nor in the
BDM. In case 4 the updated values in \cite{Dai} are 
$d_{\text{SCFA}}^+=n-w_t$  and $w_t^+ = n-d_{\text{SCFA}}$, 
thus our BDM must set:\\
$$
d_{\text{BDM}}^+ 
\stackrel{(**)}{=} 
d_{\text{SCFA}}^+ - \left\lceil\frac{n\cdot M}{M+1}\right\rceil
\stackrel{\cite{Dai}}{=} 
(n-w_t) - \left\lceil\frac{n\cdot M}{M+1}\right\rceil
\stackrel{***}{=} 
\left\lfloor\frac{n}{M+1}\right\rfloor + b_t 
- \left\lfloor\frac{n}{M+1}\right\rfloor = b_t
$$
and
$$\hspace*{- 1cm}\ds
b^+_t \stackrel{(***)}{=} 
\left\lfloor\frac{n}{M+1}\right\rfloor - w_t^+
\stackrel{\cite{Dai}}{=} 
\left\lfloor\frac{n}{M+1}\right\rfloor - (n - d_{\text{SCFA}})
\stackrel{(**)}{=} 
-\left\lceil\frac{n\cdot M}{M+1}\right\rceil+ 
d_{\text{BDM}} +\left\lceil\frac{n\cdot M}{M+1}\right\rceil
=d_{\text{BDM}},
$$
that is interchange of $d_\text{BDM}$ with $b_t$, as takes place
in a  discharge.

Finally, our transition probability over all $M$ subcycles of $D, I$
or -- is the product 
$\ds
\left(\frac{q-1}{q}\right)^{\#D}\cdot
\left(\frac{1}{q}\right)^{\#I}\cdot
\left(\frac{q}{q}\right)^{\#-}$
(where $\#D+\#I+\#-=M$), corresponding to 
$(q-1)^{\#D}1^{\#I}q^{\#-}$ different $M$--tuples of symbols in row
$n$ of the $M$ formal power series ($\#D$ times $\delta\neq 0$,
$\#I$ times $\delta= 0$, $\#-$ times {\it any} symbol from $\ff_q$).

\subsection*{III. Numerical Results about\\
Multidimensional Linear and Jump Complexity}

We have an infinite model and finite aproximations that simulate the
behaviour of the multidimensional continued fraction expansion
algorithm:  The drain $d$
corresponds to the linear complexity deviation $d=deg(v) -
\left\lceil\frac{n\cdot M}{M+1}\right\rceil$, whereas each ``D'' in a
transition corresponds to a jump by a height $b_t-d$.

We start at time 0 with a probability distribution of $pr(0,\dots,0;0)=1$, 
zero everywhere else, and run the state transition
matrix until reaching the stationary equilibrium.

\subsubsection*{a) Linear complexity deviation}

The average linear complexity deviation in level $T$ is 
(${\cal  P}(M,q)$ as in I.d):
$$\overline d(M,T) = \frac
{\sum_{s\in Q_T} q^{-K(s)} \cdot d(s)}
{\sum_{s\in Q_T} q^{-K(s)}}
=\frac{\sum_{s\in Q_T} q^{-K(s)} \cdot d(s)}
{{\cal  P}(M,q)},\ T=0,\dots,M.$$

Also, we have $\overline d(M)=\sum_{T=0}^M \overline d(M,T)/(M+1)$ as
average over all $T$.

$\overline d(M)$ turns out to be zero for all $M$ and $q$, another
argument for our choice of $deg(v)\approx 
\left\lceil\frac{n\cdot M}{M+1}\right\rceil$ as ``typical'' behaviour.

The probability that the degree deviation has a certain value
$d_0$, for some $T$, is $pr(d=d_0)_{(M,T)}=\left(\sum_{s\in
  Q_T,d(s)=d_0} q^{-K(s)}\right)/{\cal P}(M,q)$ and we set 
$pr(d=d_0)_{M}=\frac{1}{M+1}\sum_{T=0}^M pr(d=d_0)_{(M,T)}.$
We have $pr(d=d_0)_{M}=pr(d=-d_0)_{M}$.

The general (in $q$) formula for $\overline d(M,T)$ is 
$\overline d(1,0)= -\overline d(1,1)=q/(q+1)^2$,
$\overline d(2,0)= -\overline
d(2,2)=(q^5+q^4-q^3+q^2+q)/(q^3+1)(q^2+q+1)^2$,
$\overline d(2,1)= 0$, and in general 
$\overline d(M,T)=-\overline d(M,M-T)$ for $0\leq T\leq M$, leading to  
$\overline d(M)=0$ (all this by numerical evidence). For $q=2$, we obtain

$\ba{cllllllllll}
M
&\overline{d}(M,0)
&\overline{d}(M,1)
&\overline{d}(M,2)
&\overline{d}(M,3)
&\overline{d}(M,4)
\\
1&0.222222&-0.222222\\
2&0.312925&\hy 0&-0.312925\\
3&0.352021&\hy 0.10806&-0.10806&-0.352021\\
4&0.370890&\hy 0.163309&\hy 0&-0.163309&-0.370890\\
5&0.380275&\hy 0.191638&\hy 0.0572067&-0.0572067&-0.191638&\\
6&0.384972&\hy 0.206045&\hy 0.0868915&\hy 0&-0.0868915\\
7&0.387277&\hy 0.213270&\hy 0.1019991&\hy 0.0297877&-0.0297877\\
8&0.388441&\hy 0.216919&\hy 0.1096532&\hy 0.0450379&\hy 0\\
\ea$

$\ba{clllllllllll}
M&p(d=0)&p(d=\pm 1)&p(d=\pm 2)&p(d=\pm 3)&p(d=\pm 4)&p(d=\pm 5)\\
1&0.5&0.1875&0.046875&0.011719&0.002930&0.000732\\
2&0.55&0.19414&0.026959&0.03412&0.000427&5.34e-5\\
3&0.61920&0.176843&0.012701&0.0008006&5.0066e-5&3.1292e-6\\
\ea$

For $q=100$ (remember that our model requires only $2\leq q\in\rr$),
the values for $d(M,T)$ (and similar for all the other results)
suggest formal power series in $q^{-1}$, as such valid for any $q$
(the dots separate the powers of $q^{-1}$):

\bt{cll}
$M$&$\overline d(M,0)$&$\overline d(M,1)$\\
1&0,00.98.02.96.04.94.06.92\\
1&0,00.99.97.03.01.94.04.02.86&0\\
3&0,00.99.98.98.03.01.98.92.99&0,00.00.98.99.98.03.00.00.96.9\\
\et

For $M=1$ and 2 the closed form was already given, for $M=3$ we
obtain:

$\overline d(3,0) = 1q^{-1}+ 0q^{-2}- 1q^{-3}-2q^{-4}+ 3q^{-5}+ 2q^{-6}- 1q^{-7}-7q^{-8}\pm\dots$

$\overline d(3,1) = 0q^{-1}+ 1q^{-2}- 1q^{-3}+0q^{-4}-2q^{-5}+
3q^{-6}+0q^{-7}+1q^{-8}-3q^{-9}\pm\dots$

\subsubsection*{b) Jump complexity}
 
The jump complexity counts how many discharges occur, and with which
height $b_t-d$. Let $s_1\stackrel{t}{\longrightarrow}s_2$ with
$t\in\{I,D,-\}^M$ be some transition, where $t$ denotes the actions at
the $M$ batteries. Let $t_I,t_D,t_-$ be the respective number of
symbols $I,D,$ and $-$ in $t$, then $t$ has overall probability 
$\ds \frac{q^{-K(s_1)}}{{\cal P}(M,q)}\cdot
\frac{(q-1)^{t_D}}{q^{t_I+t_D}}$. Hence, we have an average jump
complexity per time unit of
$$\overline J(T)=
\sum_
{s_1\stackrel{t}{\rightarrow}s_2,s_1\in Q_T,s_2\in Q_{T+1}} 
t_D\cdot\frac{q^{-K(s_1)}}{{\cal P}(M,q)}\cdot
(q-1)^{t_D}q^{-(t_I+t_D)},$$
hence up to $n$ an expected average of $n\cdot \overline J := n\cdot
\frac{1}{M+1}\sum_{T=0}^M \overline J(T)$ jumps.

Also, we calculate how many jumps by height $h\in\nn$ occur on average
as:
$$\overline{JH}(h) =\frac{1}{M+1}\sum_{T=0}^M
\sum_
{\scriptsize{
\ba{c}
s_1\stackrel{t}{\rightarrow}s_2\\  
\hspace*{-1cm}s_1\in Q_T,s_2\in Q_{T+1}\hspace*{-1cm}
\ea}} 
|\{i\ |\ b_i-d=h \text{\ in\ } t\}|
\cdot\frac{q^{-K(s_1)}}{{\cal P}(M,q)}\cdot
(q-1)^{t_D}q^{-t_I-t_D}.$$

Again, we list some values and also have a closed formula for $M=1$
and~2.

\bt{ccllllll}
$q$&$M$&$\overline J$&$\overline{JH}(1)$&$\overline{JH}(2)$& 
$\overline{JH}(3)$&$\overline{JH}(4)$&$\overline{JH}(5)$\\
2&1&0.25   &0.125  &0.0625 &0.03125&0.015625&0.007813\\
 &2&0.44444&0.29167&0.10417&0.03385&0.10417 &0.003092\\
 &3&0.58929&0.45786&0.10699&0.02023&0.003456&0.000549\\
 &4&0.69333&0.59742&0.08610&0.00895&0.000794&0.000065\\
 &5&0.76613&0.70254&0.06014&0.00329&0.000149&0.000006\\
 &6&0.81633&0.77665&0.03856&0.00109&0.000025&0.000001\\
\et

\bt{ccllllll}
$q$&$M$&$\overline J$&$\overline{JH}(1)$&$\overline{JH}(2)$& 
$\overline{JH}(3)$\\
100&1&0.495&0.49005&0.0049005&0.000049005\\
&2&0.6665346534&0.66640266&0.00013197373&$\dots$\\
\et

That is for $M=1$ we have 
$\overline J = \frac{1}{2}-\frac{1}{q}+\frac{1}{2q}$ and 
$\overline{JH}(h) =
q^{-h+1}\cdot\left(\frac{1}{2}-\frac{1}{q}+\frac{1}{2q^2}\right)$,
and for $M=2$ we obtain by evaluating for several $q$: $\overline
J=\frac{2}{3}-\frac{4}{3q(q+1)}$. 
Observe that for $q\to \infty$ and any $M$, we have $\overline J =
\frac{M}{M+1}$, according to  model I.b).

\subsubsection*{Open Problems:}

1. Show {\it algebraically} that $\forall s_1,s_2\in Q_T$ we have 
$\ds\frac{pr(s_1)}{pr(s_2)} =q^{-K(s_1)+K(s_2)}$, where $K(s)$ is defined
via the number of inhibitions from $(0,\dots,0;0)$.

2. Show {\it algebraically} that $|\{s\in Q_T\ |\ K(s)=K \}=p_K(M)$
   for all $K,T,M$.

3. Give a closed form for the coefficients of all the new formal power
   series in $\zz[[q^{-1}]]$ occuring in this paper.

\subsubsection*{Conclusion}

We developed a model of multidimensional linear and jump
complexity, using a stochastic infinite state machine, which is
selfsimilar on the time axis, folding back time mod $M+1$ onto itself.

Fixing an arbitrary good accuracy level $K_0$, we obtain a {\it finite} model 
that approximates with an {\it exponentially} small (in $K_0$) error, 
using only {\it  polynomially} many states.

We derived values for linear and jump complexity of multisequences in
the average case and 
probabilities for deviations from that case.

The whole theory is valid for {\it any} $q$ (order of finite field),
{\it any} $M$ (number of sequences) and {\it any} timestep $n$, We
have numerical results for $M$ up to $8$, $n\to \infty$, and any $q$,
extending considerably the range of known results.

\newpage
{\bf Appendix} State counts for $M=3$, $T=0$,  $K=0,1,\dots,50$ (vertical) and\\ 
$d= -10,-9,\dots,9$ (horizontal), $d=0$ within dots. The last column is
$p_K(M)$.

{\scriptsize{
\bt{rrrrrrrrrrrrrrrrrrrrrrrrrrrrrrrr}
  0&    &    &    &    &    &    &    &    &    &    &.   1.&    &    &    &    &    &    &    &    &    &   1\\
  1&    &    &    &    &    &    &    &    &    &    &.   0.&   1&    &    &    &    &    &    &    &    &   1\\
  2&    &    &    &    &    &    &    &    &    &    &.   1.&   1&    &    &    &    &    &    &    &    &   2\\
  3&    &    &    &    &    &    &    &    &    &    &.   2.&   1&    &    &    &    &    &    &    &    &   3\\
  4&    &    &    &    &    &    &    &    &    &   1&.   2.&   1&    &    &    &    &    &    &    &    &   4\\
  5&    &    &    &    &    &    &    &    &    &   1&.   1.&   2&   1&    &    &    &    &    &    &    &   5\\
  6&    &    &    &    &    &    &    &    &    &   2&.   1.&   3&   1&    &    &    &    &    &    &    &   7\\
  7&    &    &    &    &    &    &    &    &    &   2&.   2.&   2&   2&    &    &    &    &    &    &    &   8\\
  8&    &    &    &    &    &    &    &    &   1&   3&.   3.&   1&   2&    &    &    &    &    &    &    &  10\\
  9&    &    &    &    &    &    &    &    &   1&   2&.   3.&   2&   3&   1&    &    &    &    &    &    &  12\\
 10&    &    &    &    &    &    &    &    &   2&   3&.   2.&   3&   3&   1&    &    &    &    &    &    &  14\\
 11&    &    &    &    &    &    &    &    &   2&   3&.   2.&   4&   3&   2&    &    &    &    &    &    &  16\\
 12&    &    &    &    &    &    &    &   1&   3&   4&.   3.&   3&   3&   2&    &    &    &    &    &    &  19\\
 13&    &    &    &    &    &    &    &   1&   3&   3&.   4.&   2&   4&   3&   1&    &    &    &    &    &  21\\
 14&    &    &    &    &    &    &    &   2&   4&   3&.   4.&   3&   4&   3&   1&    &    &    &    &    &  24\\
 15&    &    &    &    &    &    &    &   2&   4&   4&.   3.&   4&   4&   4&   2&    &    &    &    &    &  27\\
 16&    &    &    &    &    &    &   1&   3&   5&   4&.   3.&   5&   3&   4&   2&    &    &    &    &    &  30\\
 17&    &    &    &    &    &    &   1&   3&   4&   4&.   4.&   4&   4&   5&   3&   1&    &    &    &    &  33\\
 18&    &    &    &    &    &    &   2&   4&   5&   4&.   5.&   3&   5&   5&   3&   1&    &    &    &    &  37\\
 19&    &    &    &    &    &    &   2&   4&   5&   4&.   5.&   4&   5&   5&   4&   2&    &    &    &    &  40\\
 20&    &    &    &    &    &   1&   3&   5&   6&   5&.   4.&   5&   4&   5&   4&   2&    &    &    &    &  44\\
 21&    &    &    &    &    &   1&   3&   5&   5&   5&.   4.&   6&   4&   6&   5&   3&   1&    &    &    &  48\\
 22&    &    &    &    &    &   2&   4&   6&   5&   5&.   5.&   5&   5&   6&   5&   3&   1&    &    &    &  52\\
 23&    &    &    &    &    &   2&   4&   6&   5&   5&.   6.&   4&   6&   6&   6&   4&   2&    &    &    &  56\\
 24&    &    &    &    &   1&   3&   5&   7&   6&   5&.   6.&   5&   6&   5&   6&   4&   2&    &    &    &  61\\
 25&    &    &    &    &   1&   3&   5&   6&   6&   6&.   5.&   6&   5&   6&   7&   5&   3&   1&    &    &  65\\
 26&    &    &    &    &   2&   4&   6&   7&   6&   6&.   5.&   7&   5&   6&   7&   5&   3&   1&    &    &  70\\
 27&    &    &    &    &   2&   4&   6&   7&   6&   6&.   6.&   6&   6&   7&   7&   6&   4&   2&    &    &  75\\
 28&    &    &    &   1&   3&   5&   7&   8&   6&   6&.   7.&   5&   7&   6&   7&   6&   4&   2&    &    &  80\\
 29&    &    &    &   1&   3&   5&   7&   7&   6&   6&.   7.&   6&   7&   6&   8&   7&   5&   3&   1&    &  85\\
 30&    &    &    &   2&   4&   6&   8&   7&   7&   7&.   6.&   7&   6&   7&   8&   7&   5&   3&   1&    &  91\\
 31&    &    &    &   2&   4&   6&   8&   7&   7&   7&.   6.&   8&   6&   7&   8&   8&   6&   4&   2&    &  96\\
 32&    &    &   1&   3&   5&   7&   9&   8&   7&   7&.   7.&   7&   7&   7&   7&   8&   6&   4&   2&    & 102\\
 33&    &    &   1&   3&   5&   7&   8&   8&   7&   7&.   8.&   6&   8&   7&   8&   9&   7&   5&   3&   1& 108\\
 34&    &    &   2&   4&   6&   8&   9&   8&   7&   7&.   8.&   7&   8&   7&   8&   9&   7&   5&   3&   1& 114\\
 35&    &    &   2&   4&   6&   8&   9&   7&   8&   8&.   7.&   8&   7&   8&   9&   9&   8&   6&   4&   2& 120\\
 36&    &   1&   3&   5&   7&   9&  10&   8&   8&   8&.   7.&   9&   7&   8&   8&   9&   8&   6&   4&   2& 127\\
 37&    &   1&   3&   5&   7&   9&   9&   8&   8&   8&.   8.&   8&   8&   8&   8&  10&   9&   7&   5&   3& 133\\
 38&    &   2&   4&   6&   8&  10&   9&   9&   8&   8&.   9.&   7&   9&   8&   8&  10&   9&   7&   5&   3& 140\\
 39&    &   2&   4&   6&   8&  10&   9&   9&   8&   8&.   9.&   8&   9&   8&   9&  10&  10&   8&   6&   4& 147\\
 40&   1&   3&   5&   7&   9&  11&  10&   8&   9&   9&.   8.&   9&   8&   9&   9&   9&  10&   8&   6&   4& 154\\
 41&   1&   3&   5&   7&   9&  10&  10&   8&   9&   9&.   8.&  10&   8&   9&   9&  10&  11&   9&   7&   5& 161\\
 42&   2&   4&   6&   8&  10&  11&  10&   9&   9&   9&.   9.&   9&   9&   9&   9&  10&  11&   9&   7&   5& 169\\
 43&   2&   4&   6&   8&  10&  11&   9&  10&   9&   9&.  10.&   8&  10&   9&   9&  11&  11&  10&   8&   6& 176\\
 44&   3&   5&   7&   9&  11&  12&  10&  10&   9&   9&.  10.&   9&  10&   9&   9&  10&  11&  10&   8&   6& 184\\
 45&   3&   5&   7&   9&  11&  11&  10&   9&  10&  10&.   9.&  10&   9&  10&  10&  10&  12&  11&   9&   7& 192\\
 46&   4&   6&   8&  10&  12&  11&  11&   9&  10&  10&.   9.&  11&   9&  10&  10&  10&  12&  11&   9&   7& 200\\
 47&   4&   6&   8&  10&  12&  11&  10&  10&  10&  10&.  10.&  10&  10&  10&  10&  11&  12&  12&  10&   8& 208\\
 48&   5&   7&   9&  11&  13&  12&  10&  11&  10&  10&.  11.&   9&  11&  10&  10&  11&  11&  12&  10&   8& 217\\
 49&   5&   7&   9&  11&  12&  12&  10&  11&  10&  10&.  11.&  10&  11&  10&  10&  11&  12&  13&  11&   9& 225\\
 50&   6&   8&  10&  12&  13&  12&  11&  10&  11&  11&.  10.&  11&  10&  11&  11&  10&  12&  13&  11&   9& 234\\

\et
}}

\begin{thebibliography}{9}
\bibitem{CV} 
M\'onica del Pilar Canales Chac\'on, Michael~Vielhaber, 
{\it Isometries of binary formal power series and 
their shift commutators}, Electronic Colloquium on Computational Complexity,
ECCC TR04--057. {\tt eccc.hpi-web.de/eccc-reports/2004/TR04-57}

\bibitem{Dai}
Z.~Dai, X.~Feng,  {\it Multi-Continued fraction Algorithm and
Generalized \linebreak B--M Algorithm over $\ff_2$}, in \cite{Seta04}.

\bibitem{NV}
H.~Niederreiter, M.~Vielhaber, {\it Linear complexity profiles:
Hausdorff dimensions for almost perfect profiles and measures for general
profiles}, J.~of Complexity 13, No.~3, 353 -- 383, 1997.

\bibitem{NVsim}
H.~Niederreiter, M.~Vielhaber, {\it Simultaneous shifted continued
  fraction expansions in quadratic time}, {\bf AAECC 9, (2)}, 125 --
138, 1998.

\bibitem{VSETA}
M.~Vielhaber, {\it A Unified View on Sequence Complexity Measyures as
  Isometries}, in \cite{Seta04}.

\bibitem{V}
M.~Vielhaber, {\it Continued Fraction Expansion as Isometry: The Law
  of the Iterated logarithm for Linear, Jump, and $2$--Adic
  Complexity} Submitted in revised form to IEEE Trans.~Inform.~Th.\ \ 
Preprint: {\tt arxiv.org/CS/0511089}

\bibitem{Seta04}
Pre--Proceedings SETA '04 2004 International Conference on Sequences and 
Their Applications, October 24 -- 28, 2004, Seoul, Korea, and LNCS
3468, Springer, 2005.

\bibitem{INRIA}
{\tt algo.inria.fr/encyclopedia/formulaire.html}, ECS 352--359.

\bibitem{NJAS}
{\tt www.research.att/\~{}njas/sequences}, A001399$\dots$A001401.
\end{thebibliography}
\end{document}